\documentclass[a4paper]{jpconf}
\usepackage{graphicx}
\newcommand{\ba}{\begin{eqnarray}}
\newcommand{\ea}{\end{eqnarray}}

\begin{document}

\title{Triangular symmetry in cluster nuclei}

\author{R. Bijker and A.H. Santana-Vald\'es}
\address{Instituto de Ciencias Nucleares, 
Universidad Nacional Aut\'onoma de M\'exico, 
A.P. 70-543, 04510 Ciudad de M\'exico, M\'exico}
\ead{bijker@nucleares.unam.mx}

\begin{abstract}
In this contribution, we present evidence for the occurrence of triangular symmetry in cluster 
nuclei. We discuss the structure of rotational bands for $3\alpha$ and $3\alpha+1$ configurations 
with triangular ${\cal D}_{3h}$ symmetry by exploiting the double group ${\cal D}'_{3h}$, 
and study the application to $^{12}$C and ${13}$C. The structure of rotational bands can 
be used as a fingerprint of the underlying geometric configuration of $\alpha$ particles. 
\end{abstract}

\section{Introduction}

The study of cluster degrees of freedom in light nuclei, in particular nuclei with $A=4k$ and $4k+x$ 
nucleons goes back to early work by Wheeler \cite{wheeler}, and Hafstad and Teller \cite{Teller}, 
followed by later work by Brink \cite{Brink1,Brink2} and Robson \cite{Robson1,Robson2}. Recently, 
there has been a lot of renewed interest in the structure of $\alpha$-cluster nuclei, especially for 
the nucleus $^{12}$C \cite{FreerFynbo}. The measurement of new rotational excitations of the ground state \cite{Freer2007,Kirsebom,Marin} and of the Hoyle state \cite{Itoh,Freer2012,Gai,Freer2011} has stimulated 
a large theoretical effort to understand the structure of $^{12}$C (for a review see {\it e.g.} Refs.~\cite{FreerFynbo,Schuck,Freer}). In addition, there are measurements of many new states in 
$^{13}$C \cite{Ivano}. 

In this contribution, we present evidence for triangular symmetry in both even- and odd-cluster 
nuclei, $^{12}$C and $^{13}$C, respectively.

\section{Triangular symmetry in $^{12}$C}

The symmetry group of the equilateral triangle is the point group ${\cal D}_{3h}$. The properties 
of ${\cal D}_{3h}$ and the double group ${\cal D}'_{3h}$ are well-known in molecular physics 
\cite{Herzberg} and crystal physics \cite{Koster}. Here we summarize the results relevant for 
applications to $\alpha$-cluster nuclei in nuclear physics \cite{Bijker2000,Bijker2002,Bijker2016,PPNP}.  

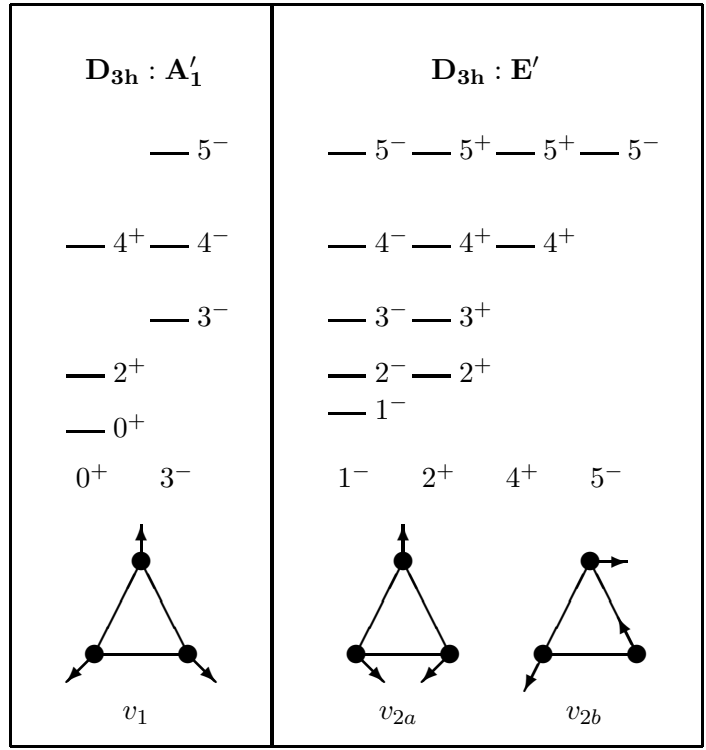
\begin{figure}
\centering
\setlength{\unitlength}{0.7pt} 
\begin{picture}(370,400)(-20,0)
\thicklines
\put (-20,  0) {\line(1,0){370}}
\put (-20,400) {\line(1,0){370}}
\put (-20,  0) {\line(0,1){400}}
\put (120,  0) {\line(0,1){400}}
\put (350,  0) {\line(0,1){400}}
\put ( 25, 50) {\circle*{10}}
\put ( 75, 50) {\circle*{10}}
\put ( 50,100) {\circle*{10}}
\put ( 25, 50) {\line ( 1,0){50}}
\put ( 25, 50) {\line ( 1,2){25}}
\put ( 75, 50) {\line (-1,2){25}}
\put ( 50,100) {\vector( 0, 1){20}}
\put ( 25, 50) {\vector(-1,-1){15}}
\put ( 75, 50) {\vector( 1,-1){15}}
\put ( 10,170) {\line(1,0){20}}
\put ( 10,200) {\line(1,0){20}}
\put ( 10,270) {\line(1,0){20}}
\put ( 35,167) {$0^+$}
\put ( 35,197) {$2^+$}
\put ( 35,267) {$4^+$}
\put ( 55,230) {\line(1,0){20}}
\put ( 55,270) {\line(1,0){20}}
\put ( 55,320) {\line(1,0){20}}
\put ( 80,227) {$3^-$}
\put ( 80,267) {$4^-$}
\put ( 80,317) {$5^-$}
\put ( 15,140) {$0^+$}
\put ( 60,140) {$3^-$}
\put ( 20,360) {$\bf D_{3h}: A'_1$}
\put (165, 50) {\circle*{10}}
\put (215, 50) {\circle*{10}}
\put (190,100) {\circle*{10}}
\put (165, 50) {\line ( 1,0){50}}
\put (165, 50) {\line ( 1,2){25}}
\put (215, 50) {\line (-1,2){25}}
\put (190,100) {\vector( 0, 1){20}}
\put (165, 50) {\vector( 1,-1){15}}
\put (215, 50) {\vector(-1,-1){15}}
\put (265, 50) {\circle*{10}}
\put (315, 50) {\circle*{10}}
\put (290,100) {\circle*{10}}
\put (265, 50) {\line ( 1,0){50}}
\put (265, 50) {\line ( 1,2){25}}
\put (315, 50) {\line (-1,2){25}}
\put (290,100) {\vector( 1, 0){20}}
\put (265, 50) {\vector(-1,-2){10}}
\put (315, 50) {\vector(-1, 2){10}}
\put ( 40, 15) {$v_1$}
\put (177, 15) {$v_{2a}$}
\put (277, 15) {$v_{2b}$}
\put (150, 180) {\line(1,0){20}}
\put (150, 200) {\line(1,0){20}}
\put (150, 230) {\line(1,0){20}}
\put (150, 270) {\line(1,0){20}}
\put (150, 320) {\line(1,0){20}}
\put (175, 177) {$1^-$}
\put (175, 197) {$2^-$}
\put (175, 227) {$3^-$}
\put (175, 267) {$4^-$}
\put (175, 317) {$5^-$}
\put (195, 200) {\line(1,0){20}}
\put (195, 230) {\line(1,0){20}}
\put (195, 270) {\line(1,0){20}}
\put (195, 320) {\line(1,0){20}}
\put (220, 197) {$2^+$}
\put (220, 227) {$3^+$}
\put (220, 267) {$4^+$}
\put (220, 317) {$5^+$}
\put (240, 270) {\line(1,0){20}}
\put (240, 320) {\line(1,0){20}}
\put (265, 267) {$4^+$}
\put (265, 317) {$5^+$}
\put (285, 320) {\line(1,0){20}}
\put (310, 317) {$5^-$}
\put (155, 140) {$1^-$}
\put (200, 140) {$2^+$}
\put (245, 140) {$4^+$}
\put (290, 140) {$5^-$}
\put (205,360) {$\bf D_{3h}: E'$}
\end{picture}
\caption{Structure of rotational bands for a triangular configuration of $\alpha$ particles 
in even-cluster nuclei with $A'_1$ (left) and $E'$ symmetry (right). 
Each rotational band is labeled by $K^P$.} 
\label{roteven}
\end{figure}

For even-cluster nuclei the states can be labeled by $| \Omega,K,L \rangle$ where $\Omega$ labels 
the tensor (or bosonic) representations of the $D_{3h}$ triangular symmetry, and $K$ and $L$ are 
integers representing the projection $K$ of the angular momentum $L$ on the symmetry axis
\begin{equation}
\begin{array}{lll}
\Omega = A_{1}^{\prime}: &\hspace{0.5cm}& K^P=0^+, 3^-, 6^+, \ldots, \\ 
\Omega = E^{\prime}: && K^P=1^-, 2^+, 4^+, 5^-, \ldots, 
\end{array}
\label{even}
\end{equation}
The angular momentum content of each $K$ band is given by 
$L=0,2,4,\ldots,$ for $K=0$ and $L=K,K+1,K+2,\ldots,$ for $K>0$. . 
The rotational structure depends on the $D_{3h}$ point group symmetry of the equilateral triangle 
configuration and is summarized in Fig.~\ref{roteven}. 

\begin{figure}[h]
\centering
\includegraphics[width=4in]{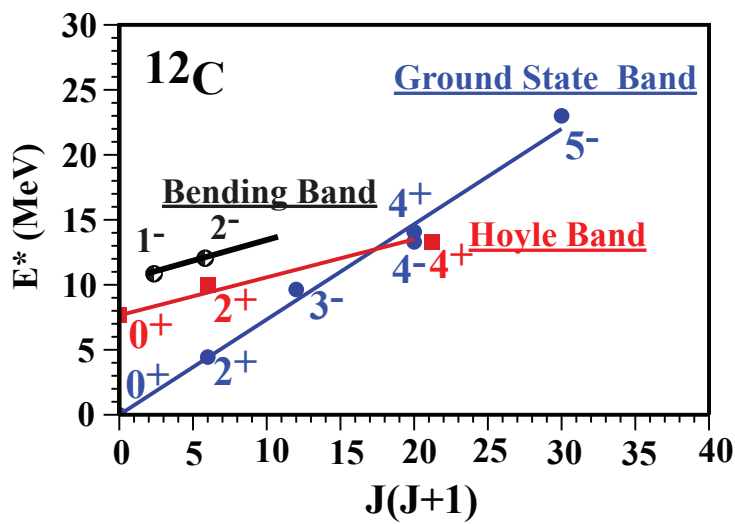} 
\caption[Rotational bands in $^{12}$C]
{Rotational bands in $^{12}$C \cite{Marin}.}
\label{bands3}
\end{figure}

The band with $A'_1$ symmetry 
is characterized by a rotational sequence involving both positive and negative parity states, 
$L^P=0^+$, $2^+$, $3^-$, $4^{\pm}$, $5^-$, $\ldots$, all of which have been observed in the 
ground-state band of $^{12}$C. The so-called Hoyle band has the same structure, 
but so far only the positive parity states have been observed. There is some evidence for 
an excited band with $E'$ symmetry. The rotational bands in $^{12}$C are shown in Fig.~\ref{bands3}. 

\section{Triangular symmetry in $^{13}$C}

For odd-cluster nuclei the states can be labeled by $| \Omega,K,J \rangle$ where $\Omega$ now labels 
the spinor (or fermionic) representations of the double group $D'_{3h}$, and $K$ and $J$ are half 
integers representing the projection $K$ of the total angular momentum $J$ on the symmetry axis 
\cite{Bijker2019}
\begin{equation}
\begin{array}{lllll}
\Omega = E_{1/2}^{(+)}: &\hspace{0.5cm}& K^P=\frac{1}{2}^+, \frac{5}{2}^-, \frac{7}{2}^-, 
\frac{11}{2}^+, \frac{13}{2}^+, \ldots, \\
\Omega = E_{1/2}^{(-)}: && K^P=\frac{1}{2}^-, \frac{5}{2}^+, \frac{7}{2}^+, 
\frac{11}{2}^-, \frac{13}{2}^-, \ldots, \\
\Omega = E_{3/2}: && K^P=\frac{3}{2}^{\pm}, \frac{9}{2}^{\pm}, \frac{15}{2}^{\pm}, \ldots, 
\end{array}
\label{odd}
\end{equation}
The angular momenta of each $K$ band are given by $J=K,K+1,K+2,\ldots$. 
The angular momentum structure of each one of the representations of $D'_{3h}$ is shown in Fig.~\ref{rotbands}.  

\begin{figure}[h]
\centering
\setlength{\unitlength}{0.5pt}
\begin{picture}(630,450)(160,-30)
\thicklines
\put (160,-30) {\line(1,0){630}}
\put (160,420) {\line(1,0){630}}
\put (160,-30) {\line(0,1){450}}
\put (390,-30) {\line(0,1){450}}
\put (620,-30) {\line(0,1){450}}
\put (790,-30) {\line(0,1){450}}
\put (230,360) {$\bf D'_{3h}: E_{1/2}^{(-)}$}
\put (460,360) {$\bf D'_{3h}: E_{1/2}^{(+)}$}
\put (660,360) {$\bf D'_{3h}: E_{3/2}$}
\put (190, 60) {\line(1,0){20}}
\put (190, 90) {\line(1,0){20}}
\put (190,140) {\line(1,0){20}}
\put (190,210) {\line(1,0){20}}
\put (190,300) {\line(1,0){20}}
\put (250,140) {\line(1,0){20}}
\put (250,210) {\line(1,0){20}}
\put (250,300) {\line(1,0){20}}
\put (310,210) {\line(1,0){20}}
\put (310,300) {\line(1,0){20}}
\put (200, 10) {$\bf \frac{1}{2}^-$}
\put (215, 55) {$\bf \frac{1}{2}^-$}
\put (215, 85) {$\bf \frac{3}{2}^-$}
\put (215,135) {$\bf \frac{5}{2}^-$}
\put (215,205) {$\bf \frac{7}{2}^-$}
\put (215,295) {$\bf \frac{9}{2}^-$}
\put (260, 10) {$\bf \frac{5}{2}^+$}
\put (275,135) {$\bf \frac{5}{2}^+$}
\put (275,205) {$\bf \frac{7}{2}^+$}
\put (275,295) {$\bf \frac{9}{2}^+$}
\put (320, 10) {$\bf \frac{7}{2}^+$}
\put (335,205) {$\bf \frac{7}{2}^+$}
\put (335,295) {$\bf \frac{9}{2}^+$}
\put (420, 60) {\line(1,0){20}}
\put (420, 90) {\line(1,0){20}}
\put (420,140) {\line(1,0){20}}
\put (420,210) {\line(1,0){20}}
\put (420,300) {\line(1,0){20}}
\put (480,140) {\line(1,0){20}}
\put (480,210) {\line(1,0){20}}
\put (480,300) {\line(1,0){20}}
\put (540,210) {\line(1,0){20}}
\put (540,300) {\line(1,0){20}}
\put (430, 10) {$\bf \frac{1}{2}^+$}
\put (445, 55) {$\bf \frac{1}{2}^+$}
\put (445, 85) {$\bf \frac{3}{2}^+$}
\put (445,135) {$\bf \frac{5}{2}^+$}
\put (445,205) {$\bf \frac{7}{2}^+$}
\put (445,295) {$\bf \frac{9}{2}^+$}
\put (490, 10) {$\bf \frac{5}{2}^-$}
\put (505,135) {$\bf \frac{5}{2}^-$}
\put (505,205) {$\bf \frac{7}{2}^-$}
\put (505,295) {$\bf \frac{9}{2}^-$}
\put (550, 10) {$\bf \frac{7}{2}^-$}
\put (565,205) {$\bf \frac{7}{2}^-$}
\put (565,295) {$\bf \frac{9}{2}^-$}
\put (650, 90) {\line(1,0){20}}
\put (650,140) {\line(1,0){20}}
\put (650,210) {\line(1,0){20}}
\put (650,300) {\line(1,0){20}}
\put (710,300) {\line(1,0){20}}
\put (660, 10) {$\bf \frac{3}{2}^\pm$}
\put (675, 85) {$\bf \frac{3}{2}^\pm$}
\put (675,135) {$\bf \frac{5}{2}^\pm$}
\put (675,205) {$\bf \frac{7}{2}^\pm$}
\put (675,295) {$\bf \frac{9}{2}^\pm$}
\put (720, 10) {$\bf \frac{9}{2}^\pm$}
\put (735,295) {$\bf \frac{9}{2}^\pm$}
\end{picture}
\caption{Structure of rotational bands for a triangular configuration of $\alpha$ particles 
in odd-cluster nuclei with $E_{1/2}^{(-)}$, $E_{1/2}^{(+)}$ and $E_{3/2}$ symmetry. 
Each rotational band is labeled by $K^P$.}
\label{rotbands}
\end{figure}

\section{The cluster shell model}

The structure of single-particle levels moving in the deformed field of the cluster potential 
has been studied recently in the context of the Cluster Shell Model (CSM) \cite{PPNP,CSM,Valeria}. 
The CSM combines cluster and single-particle degrees of freedom, and is very similar in spirit as 
the Nilsson model \cite{Nilsson}, but in the CSM the odd nucleon moves in the deformed field generated 
by the (collective) cluster degrees of freedom. For a cluster potential with triangular symmetry the 
single-particles levels of a neutron split according to the irreducible representations of the 
double group ${\cal D}'_{3h}$, $\Omega=E_{1/2}^{(-)}$, $E_{1/2}^{(+)}$ and $E_{3/2}$, each of which is 
doubly degenerate. The resolution of single-particle levels into representations 
of $D'_{3h}$ is shown in Table~\ref{splitting}. 

\begin{table}[h]
\centering
\caption{Resolution of single-particle levels into irreducible representations of $D'_{3h}$. 
Each $E$ level is doubly degenerate.}
\label{splitting}
\vspace{10pt}
\begin{tabular}{cccccccc}
\hline
\hline
\noalign{\smallskip}
& $E_{1/2}^{(+)}$ & $E_{1/2}^{(-)}$ & $E_{3/2}$ &
& $E_{1/2}^{(+)}$ & $E_{1/2}^{(-)}$ & $E_{3/2}$ \\
\noalign{\smallskip}
\hline
\noalign{\smallskip}
$1s_{1/2}$ & 1 & 0 & 0 & $2s_{1/2}$ & 1 & 0 & 0 \\
\noalign{\smallskip}
$1p_{1/2}$ & 0 & 1 & 0 & $1d_{3/2}$ & 1 & 0 & 1 \\
\noalign{\smallskip}
$1p_{3/2}$ & 0 & 1 & 1 & $1d_{5/2}$ & 1 & 1 & 1 \\
\noalign{\smallskip}
\hline
\end{tabular}
\end{table}

A study of the neutron levels for $^{12}$C shows that the first six neutrons occupy the intrinsic states 
with $\Omega=E_{1/2}^{(+)}$ (arising from the $1s_{1/2}$ orbit), $E_{3/2}$ and $E_{1/2}^{(-)}$ (from the 
$1p_{3/2}$ orbit), so that the extra neutron in $^{13}$C occupies the intrinsic state with $E_{1/2}^{(-)}$ 
(from the $1p_{1/2}$ orbit), followed by $E_{1/2}^{(+)}$, $E_{1/2}^{(-)}$ and $E_{3/2}$ associated with 
the orbits from the $s$-$d$ shell \cite{PPNP,CSM,Cocoyoc}. 

The rotational energy spectra can be analyzed with 
\ba
E_{\rm rot}(\Omega,K,J) &=& \varepsilon_{\Omega} + A_{\Omega} \left[ J(J+1) + b_{\Omega} K^{2}  
+ a_{\Omega} (-1)^{J+1/2}(J+1/2) \delta_{K,1/2} \right] ~,
\label{erot3}
\ea
where $\varepsilon_{\Omega}$ is the intrinsic energy, $A_{\Omega} = \hbar^{2}/2{\cal I}$  
the inertial parameter, $b_{\Omega}$ a Coriolis term, and $a_{\Omega}$ the decoupling parameter. 
The latter term applies only to representations $\Omega = E_{1/2}^{(\pm)}$ and $K^P=1/2^{\pm}$.
Eq.~(\ref{erot3}) is the same energy formula as used by Nuhn in Ref.~\cite{Nuhn} in a description of 
the bandheads of $^{13}$C in the context of a two-center shell model description of the system 
$^{13}\mbox{C} + ^{16}\mbox{O} \rightarrow ^{29}$Si. Nuhn used an axially symmetric potential, 
in contrast to a cluster potential with triangular $D_{3h}$ symmetry in the CSM. As a consequence, 
in the CSM the $K^P=1/2^-$ and $5/2^+$ bands belong to the same configuration $\Omega=E_{1/2}^{(-)}$
(see Eq.~(\ref{odd}) and Fig.~\ref{rotbands}), whereas in the axially symmetric case they 
represent separate rotational bands. 

\begin{figure}[h]
\centering
\begin{minipage}{0.4\linewidth}
\includegraphics[width=3in]{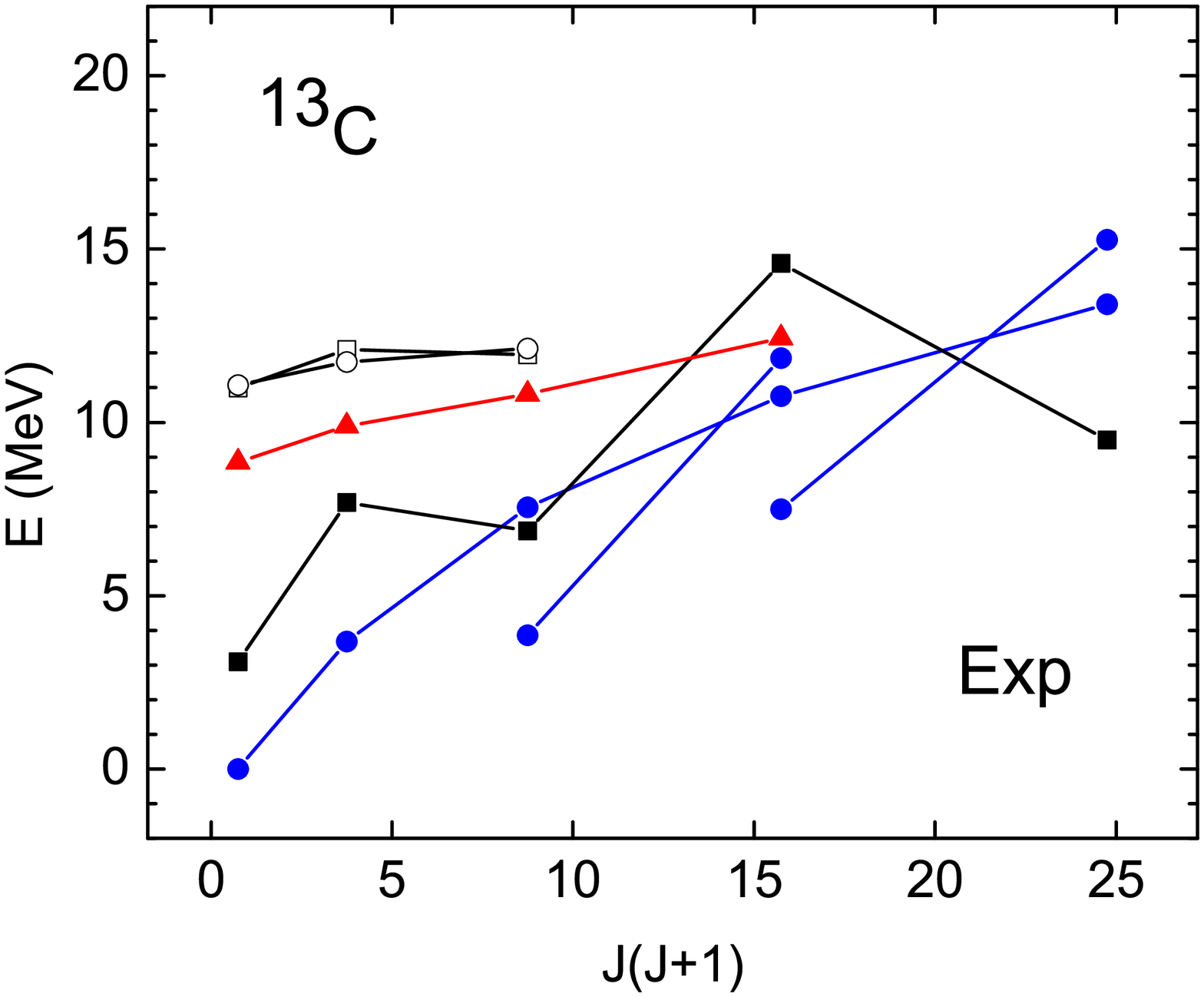}
\end{minipage}
\begin{minipage}{0.5\linewidth}
\includegraphics[width=3in]{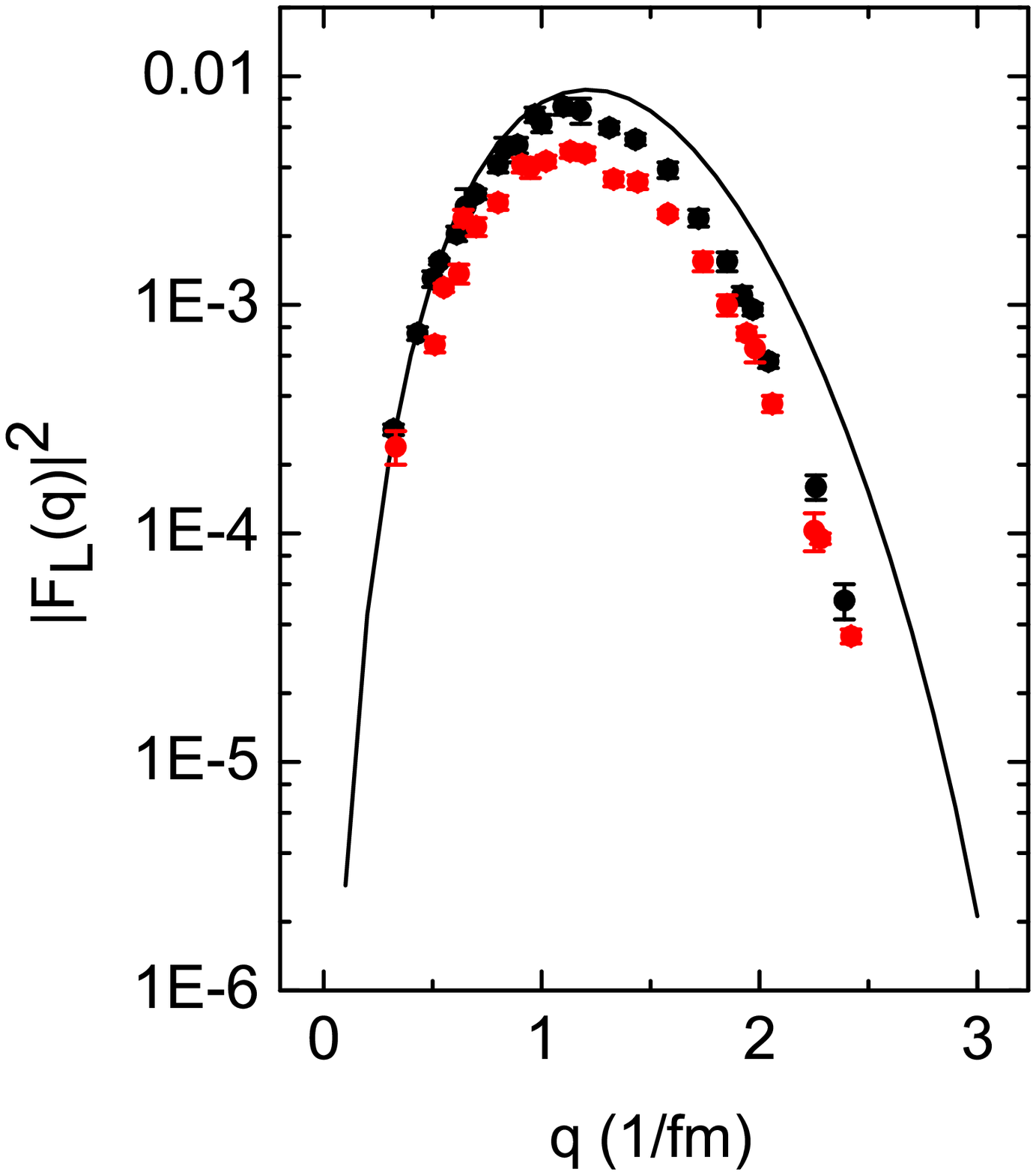}
\end{minipage}
\caption{Left: Rotational bands in $^{13}$C \cite{Bijker2019}. Right:  
Comparison between calculated and experimental \cite{millener2} longitudinal 
$E2$ form factors for the ground-state band of $^{13}$C, $F(q;1/2^-_1 \rightarrow 5/2^-_1)$
(black) and $F(q;1/2^-_1 \rightarrow 3/2^-_1)$ (red).}
\label{C13}
\end{figure}

Fig.~\ref{C13} shows the rotational bands of $^{13}$C. The ground-state band has $K^{P}=1/2^-$ 
and is assigned to the representation $\Omega=E_{1/2}^{(-)}$ of $D_{3h}^{\prime}$ (blue lines and filled 
circles) arising from the coupling of the ground-state band in $^{12}$C to the intrinsic state with 
$E_{1/2}^{(-)}$. According to Eq.~(\ref{odd}), this representation contains also $K^{P}=5/2^{+}$ 
and $7/2^{+}$ bands, both of which appear to have been observed. In the shell model, positive parity 
states are expected to occur at much higher energies since they come from the $s$-$d$ shell. 
The first excited rotational band has $K^{P}=1/2^{+}$ which can be assigned to $\Omega=E_{1/2}^{(+)}$ 
(black line and filled squares) arising from the coupling of the ground-state band in $^{12}$C to the 
excited intrinsic state with $E_{1/2}^{(+)}$. In contrast to the ground-state band, this excited band 
has a large decoupling parameter. In addition, Fig.~\ref{C13} shows evidence for the occurrence of 
a rotational band at an energy slightly higher than that of the Hoyle state in $^{12}$C which is 
interpreted as the coupling of the Hoyle band in $^{12}$C to the ground-state intrinsic state 
$E_{1/2}^{(-)}$ (red line and filled triangles). 

Further evidence for the occurrence of $D_{3h}^{\prime }$ symmetry in $^{13}$C is provided by 
electromagnetic transition rates and form factors. As an example, we show in Fig.~\ref{C13} 
the longitudinal $E2$ form factors of the states $5/2_1^-$ and $3/2_1^-$ of the ground-state 
rotational band. The two form factors are predicted to have identical shapes 
\ba
F(q;1/2^-_1 \rightarrow 5/2^-_1) \;=\; F(q;1/2^-_1 \rightarrow 3/2^-_1) ~,
\ea
and identical $B(E2;\uparrow)$ values: 9.6 W.U. This is verified to a very good approximation.

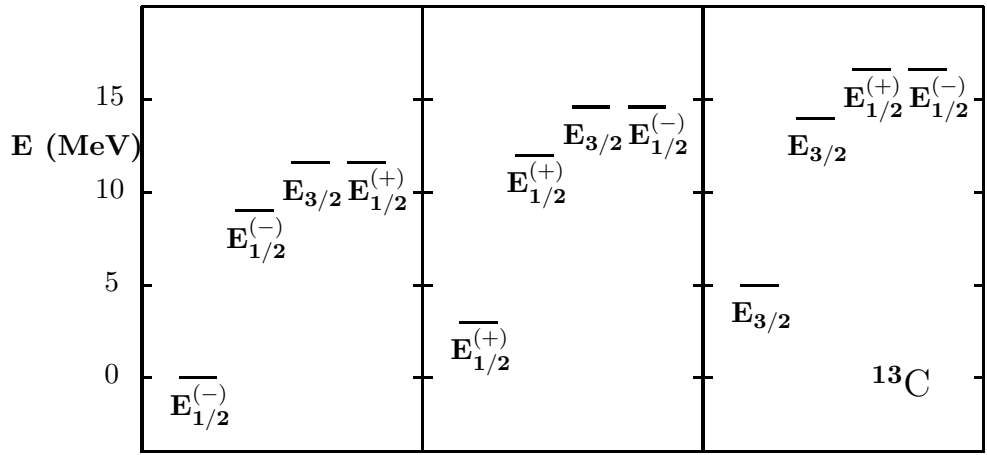
\begin{figure}[h]
\centering
\setlength{\unitlength}{0.7pt}
\begin{picture}(520,240)(-70,30)
\thicklines
\put (  0, 30) {\line(1,0){450}}
\put (  0,270) {\line(1,0){450}}
\put (  0, 30) {\line(0,1){240}}
\put (150, 30) {\line(0,1){240}}
\put (300, 30) {\line(0,1){240}}
\put (450, 30) {\line(0,1){240}}
\put (  0, 70) {\line(1,0){5}}
\put (  0,120) {\line(1,0){5}}
\put (  0,170) {\line(1,0){5}}
\put (  0,220) {\line(1,0){5}}
\put (-20, 67) {0}
\put (-20,117) {5}
\put (-25,167) {10}
\put (-25,217) {15}
\put (-70,190) {\bf E (MeV)}
\put (145, 70) {\line(1,0){10}}
\put (145,120) {\line(1,0){10}}
\put (145,170) {\line(1,0){10}}
\put (145,220) {\line(1,0){10}}
\put (295, 70) {\line(1,0){10}}
\put (295,120) {\line(1,0){10}}
\put (295,170) {\line(1,0){10}}
\put (295,220) {\line(1,0){10}}
\put (445, 70) {\line(1,0){5}}
\put (445,120) {\line(1,0){5}}
\put (445,170) {\line(1,0){5}}
\put (445,220) {\line(1,0){5}}

\put ( 20, 70) {\line(1,0){20}}
\put ( 15, 50) {$\bf E_{1/2}^{(-)}$}
\put ( 50,160) {\line(1,0){20}}
\put ( 45,140) {$\bf E_{1/2}^{(-)}$}
\put ( 80,186) {\line(1,0){20}}
\put ( 75,166) {$\bf E_{3/2}$}
\put (110,186) {\line(1,0){20}}
\put (110,166) {$\bf E_{1/2}^{(+)}$}

\put (170,100) {\line(1,0){20}}
\put (165, 80) {$\bf E_{1/2}^{(+)}$}
\put (200,190) {\line(1,0){20}}
\put (195,170) {$\bf E_{1/2}^{(+)}$}
\put (230,216) {\line(1,0){20}}
\put (225,196) {$\bf E_{3/2}$}
\put (260,216) {\line(1,0){20}}
\put (260,196) {$\bf E_{1/2}^{(-)}$}

\put (320,120) {\line(1,0){20}}
\put (315,100) {$\bf E_{3/2}$}
\put (350,210) {\line(1,0){20}}
\put (345,190) {$\bf E_{3/2}$}
\put (380,236) {\line(1,0){20}}
\put (375,216) {$\bf E_{1/2}^{(+)}$}
\put (410,236) {\line(1,0){20}}
\put (410,216) {$\bf E_{1/2}^{(-)}$}

\put (390, 60) {\large $\bf ^{13}$C}
\end{picture}
\caption{Expected vibrational spectrum of $^{13}$C for the coupling of a single-particle 
level with $E_{1/2}^{(-)}$ (left), $E_{1/2}^{(-)}$ (middle) and $E_{3/2}$ symmetry (right) 
to the ground-state band, the Hoyle band and the bending band in $^{12}$C (see Fig.~\ref{bands3}).}
\label{vib}
\end{figure}

Finally, in Fig.~\ref{vib} we show the expected structure of the vibrational spectrum of $^{13}$C. 

\section{Summary and conclusions}

\begin{figure}[h]
\centering
\includegraphics[width=0.35\linewidth]{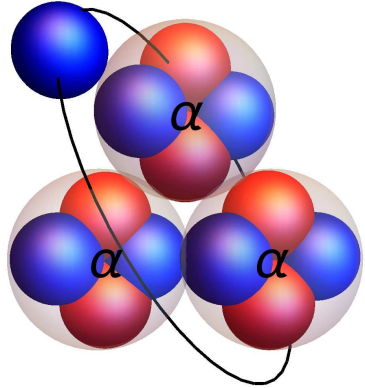}
\caption{Molecular-like picture of $^{13}$C.}
\label{3alphan}
\end{figure}

In this contribution, we presented a discussion of triangular symmetry in cluster nuclei 
and studied the application to the even- and odd-cluster nuclei, $^{12}$C and $^{13}$C.
A combined analysis of the rotation-vibration spectra and electromagnetic transition rates 
and form factors provides strong evidence for the occurrence of triangular symmetry in these nuclei.  
A characteristic feature of the triangular symmetry is the appearance of rotational 
bands consisting of both positive and negative parity states. As a consequence of the symmetry 
the form factors to the first excited state with $J^P=3/2^-$ and $5/2^-$ are predicted to have 
the same shape and $B(E2\uparrow)$ values. In addition, the quadrupole and octupole 
transitions in $^{12}$C and $^{13}$C are strongly correlated \cite{Bijker2019}. 
The good agreement between theory and experiment supports the interpretation of the nucleus 
$^{13}$C as a system of three $\alpha$-particles in a triangular configuration plus an additional 
neutron moving in the deformed field generated by the cluster (see Fig.~\ref{3alphan}). 

\ack
This work was supported in part by grant IN109017 from DGAPA-UNAM, Mexico.

\end{document}